\documentstyle[preprint,prb,eqsecnum,aps]{revtex}

\begin{document} \draft

\title{Very fast  relaxation in  polycarbonate glass}
\author{L. Saviot, E. Duval}
\address{
        Laboratoire de Physico--Chimie des Mat\'{e}riaux
        Luminescents,  UMR-CNRS 5620, Universit\'{e} Lyon I ,
				69622 Villeurbanne Cedex, France
        }
\author{J. F. Jal}
\address{D\'{e}partement de Physique des Mat\'{e}riaux, UMR-CNRS 5586,
				Universit\'{e} Lyon I, 69622 Villeurbanne
Cedex, France}
\author{A. J. Dianoux}
\address{Institut  Laue-Langevin, F38042 Grenoble Cedex, France}
\date{\today}
\maketitle

\begin{abstract}
Relaxations in amorphous  bis-phenol A polycarbonate are studied by neutron
scattering, as a function of temperature below the glass transition. Two
different processes are observed. One is very fast, with a characteristic
time  ($\sim 0.3\, ps$), that is independent of temperature and momentum
transfer. Conversely the other is slower, with a time, which is dependent
on temperature and momentum transfer. The very fast localized anharmonic
motion is interpreted by the overdamping of low-frequency vibrational
modes, by nearby dynamic holes. The slower relaxation is  thermally
activated and momentum transfer dependent. It corresponds to molecular
group motions and possibly to the short-time regime of the segmental
relaxation.
\end{abstract}


\section{INTRODUCTION}

Anharmonic motions are observed in glasses, and in particular in amorphous
polymers, by inelastic neutron scattering and inelastic light scattering
down to temperatures much lower than the temperature of glass transition,
$T_g$. These non-harmonic motions appear in the non-linear temperature
dependence of the mean square displacement, and are responsible for the
quasielastic scattering. In the case of amorphous poly(methyl methacrylate)
(PMMA) it was deduced, from neutron or light scattering measurements,  that
the high-frequency tail of the quasielastic scattering is Lorentzian, the
half width at half maximum (HWHM) of the corresponding Lorentz curve being
close to 1 meV \cite{Arr95,Sur96,Mer97}. Furthermore the HWHM was found
independent of the  momentum transfer and very weakly dependent on
temperature. On the other hand the intensity of quasielastic scattering was
shown to be strongly dependent on temperature. Such characteristics
indicate that the quasielastic scattering is due to localized and weakly
thermally activated relaxations of molecular groups. These localized
motions are released by a slower relaxation, that is thermally activated.
This last point was confirmed by the comparison of the quasielastic light
scattering (QELS) with the volume fraction of dynamic holes, which was
determined by positron annihilation lifetime spectroscopy
\cite{Duv98,Div98}. It was shown that the thermal behavior of the QELS,
that is proportional to the number of relaxing units, and of the hole
volume fraction, are identical \cite{Duv98}. From this comparison it is
clear that the temperature dependence of quasielastic intensity comes from
the one of the number of relaxing units. Similar conclusions were deduced
from measurements on amorphous bis-phenol A polycarbonate (PC)
\cite{Nov97}. \par

The study of the relaxational motions by inelastic or quasielastic neutron
scattering allows us to analyse the dynamics of relaxation in PC from about
0.1  up to 10 picoseconds. The corresponding neutron measurements are
presented in this paper. The presence of very fast localized anharmonic
motions, with a characteristic time close to 0.3 ps, was confirmed. Similar
very fast relaxations were observed in the ortho-terphenyl glass by
depolarized light scattering \cite{Pat94}, and very recently by Brillouin
scattering\cite{Mona99}. In addition another relaxation was observed, with
a time, which is dependent on temperature and momentum transfer.  In a
recent paper Colmenero and Arbe \cite{Col98} described the relaxation in
polymer glasses, such as PC, at temperature larger than a critical one,
$T_{f}$, by a model similar to the one used for the segmental
$\alpha$-relaxation in the supercooled state \cite{Col93}. According to
this model, the relaxation is Debye-like up to a critical time $\tau_{c}$,
and becomes slower and non-exponential at longer times. The slower
relaxation, that is observed in this work, could correspond partially to
the segmental one in the Debye-like regime.

In the first part of this paper the experimental results are presented. It
will be shown that the experimental neutron scattering intensities, which
are due to relaxations,  can be fitted by two exponential decays. The
experimental parameters are analyzed and discussed in the second part. \par

\section{EXPERIMENT}

The PC specimen was purchased from Bayer. The number average molecular
weight is  15 600 $mol^{-1}$ with a polydispersity index of 1.85. The glass
transition temperature is around 420 K, as determined by differential
scanning calorimetry. The thickness of the samples was equal to 0.29 mm.\par

The inelastic neutron spectra were recorded on the time-of-flight intrument
IN6 at the ILL, Grenoble. The wavelength of the incident neutrons was equal
to $5.12\,$\AA$^{-1}$ resulting in an elastic energy  resolution (FWHM) of
$80\, \mu eV$ and an elastic momentum transfer range extending from
$Q=0.22\, \AA^{-1}$ to $Q=2.06\, \AA^{-1}$. The spectra were taken in the
temperature interval 15-390 K, using a helium cryofurnace. The scattering
cross-sections were obtained after the usual standard calibrations by means
of the vanadium runs and removal of the empty-can contributions. On the
basis that PC mainly yields incoherent scattering (as was experimentally
checked), we calculated the inelastic neutron scattering intensity and,
then, the total density of states through the use of an iterative procedure
described elsewhere \cite{Fon90}. \par

\section{EXPERIMENTAL RESULTS}

To show the presence of anharmonic or relaxational motions at $T<T_g$, the
mean square displacement (MSD), $\langle u^2\rangle$, was deduced from the
Debye-Waller factor, $e^{-2W(Q)}$, where $2W(Q)=Q^2\langle u^2\rangle/3$,
which was determined by plotting the normalized elastic scattering
intensity, $S(Q, \omega =0, T)/S(Q, \omega =0, 0)$, versus $Q^2$, at
different temperatures. The logarithm of the normalized intensity is
perfectly $Q^2$ linear in the explored Q-range (0.2-1.9 \AA$^{-1}$).
However, as for other polymeric glasses,  it was observed that the
different straight lines do not converge to a same point at $Q=0$. This is
due to the competition between the elastic and the inelastic multiple
scatterings  \cite{Fric93}: the elastic multiple scattering events are
dominating at low temperature, and are gradually replaced by inelastic
scattering events with increasing temperature. This defect of elastic
scattering at $Q=0$, that is due to multiple scattering,  was very recently
confirmed  by Wuttke \cite{Wut00}. But, as shown by this author, it has a
negligible effect on the slope of straight lines in the considered Q-range
and for the thickness (0.29 mm) of the experimented samples.  The MSD,
$\langle u^2\rangle$, is plotted against T in Figure-1. It is observed that
$\langle u^2\rangle$ becomes non-linearly temperature dependent from a
temperature slightly higher than 50 K, showing the presence of anharmonic
motions from this temperature. This is confirmed by the vibrational density
of states (VDOS), which increases with temperature at low energy, from the
same temperature \cite{Sav99}. \par

Figure-2a and -2b show the Debye-Waller corrected $S(Q,\omega)$ scattering
intensities, respectively at different temperatures for $Q=1.9 \AA^{-1}$)
and at $T=390\, K$ for different Q. The inelastic scattering by harmonic
modes can be observed even at 30 K. On the other hand, the effect of
anharmonic motions is clearly present from $T = 220\, K$. Figure-2b shows
that it increases with Q. In order to obtain more informations about the
anharmonic or relaxational motions, we determined a time-dependent
scattering intensity, $S_{nh} (Q,t)$, which is independent of the harmonic
vibrations, by taking the Fourier transform of $S(Q,\omega)$, and using a
procedure similar to the one described by Colmenero et al. \cite{Col93}. In
fact $S(Q,\omega)$ is the scattering intensity at constant angle, Q being
the momentum transfer for elastic scattering. The actual momentum transfer
is larger than Q at the highest frequencies of the considered spectral
range. But this variation  is without serious consequence, since the
fastest decay will be found to be Q-independent. $S(Q,\omega)$ is the
convolution product of $S_{nh}(Q,\omega)$ by the harmonic vibration
scattering intensity, $S_{h} (Q,\omega)$, and by the instrumental
resolution, $R(Q,\omega)$:

\begin{equation}
\label{con}
S(Q,\omega)=S_{nh} (Q,\omega) \otimes S_{h} (Q,\omega) \otimes R(Q,\omega)
\end{equation}

\noindent
The product $S_{h} (Q,\omega) \otimes R(Q,\omega)$ at a temperature T is
given by the experimental scattering intensity at a low temperature, at
which the neutron scattering from relaxational motions is negligible,
corrected by the Bose and Debye-Waller factors  for the temperature T. It
was controlled that at 30 K no relaxation appears in the neutron scattering
(Figures 1 and 2). However one observed that there exists, from a
temperature of 100 K, a significant softening: An identical relative
decrease of the transverse sound velocity, that was observed by Brillouin
scattering, and of the Raman peak position or of the frequencies of
low-frequency vibrational modes was determined \cite{Niv97}. At $T=400K$
this relative decrease is close to 0.7. To take into account the softening
at a temperature T, $S(Q,\omega)$ at $T=30K$ was divided by the Bose factor
at this temperature; in a second step the frequencies were renormalized by
the relative decrease coefficient of the vibrational low frequencies; and
then the obtained structure factor was multiplied by the Bose factor at
temperature T . The Fourier transforms were performed through the real
one-dimensional tranform
provided by version 2 of the freely available FFTW library (Massachusetts
Institute of Technology - http://www.fftw.org/). The $S(Q,\omega)$ data
were taken up
to 10 meV before performing the fast Fourier transform in order to do it over
the same interval for all temperatures. Taking into account this spectral
interval and the instrumental resolution a fairly good confidence can be
given to the Fourier transforms from t=0 to t=10 ps. $S_{nh} (Q,t)$, at
temperature T, was then obtained by dividing the Fourier transform $S(Q,t)$
by the Fourier transform $S_{h} (Q,t).R(Q,t)$ of $S_{h} (Q,\omega) \otimes
R(Q,\omega)$ obtained in this way at temperature T . The relative intensity
$S_{nh} (Q,t)/S_{nh} (Q,0)$ is plotted against the time $t$ for different
temperatures and $Q=1.9\,\AA^{-1}$ in Figure-3, and for different momentum
transfers and $T=390 K$ in Figure-4. \par

Different mathematical expressions, having a physical meaning, were tested
to fit $S_{nh}(Q,t)$. Up to a time t=10 ps, that is limited by the
instrumental resolution, the  fits, which are unambiguously the best ones
(Figure 3 and 4), were obtained by two exponentials decays:
\begin{eqnarray}
\label{exp}
S_{nh}(Q,t)=((1-\rho)+\rho exp-t/\tau_{1})exp-t/\tau_{2}
\end{eqnarray}

\noindent
In this expression $\rho \leq 1$. It is emphasized that (\ref{exp}) is
considered to be valid only at short times ($t<\tau_{2}$). The time
$\tau_{1}$ is approximately equal to $0.3 ps$ and very weakly dependent on
temperature and momentum transfer. The fitting parameters are collected in
Table-1. The time $\tau_{2}$ decreases with $Q$  and $T$ ($\tau_{2}\approx
60 ps$ at $T=390 K$ and $Q=1.9\AA^{-1}$). It is likely that, at the
considered longest times and highest temperatures, the decays can be fitted
by a stretched exponential or a distribution of relaxation times. On the
other hand, as suggested by Novikov \cite{Nov00}, it is not impossible that
the decays, in the 0-10 ps interval, may be fitted by a peculiar
distribution of relaxation times. However in this case the analysis of the
experimental data is not so straightforward, as it is possible tofit with
different sets of parameters. \par

\section{DISCUSSION}

We will discuss principally the very fast anharmonic motion (symbolized by
VFAM), that is characterized by the sub-picosecond time $\tau_{1}$. At
first it is noticed that the non-linearly temperature dependent MSD
increases from a temperature much lower than $T_{g}$. It is likely that, at
temperatures higher than 120 K, the pure methyl-group rotation and the
$\pi-flip$  motions of phenylene contribute to this MSD.  It is important
to be sure that such molecular motions do not contribute to the VFAM. The
activation energy for the methyl-group rotation is weaker than the one for
$\pi-flip$ motion: From nuclear magnetic resonance, they are respectively
15-20 kJ/mole \cite{Schm85}, and 30-50 kJ/mole \cite{Weh87}.  The possible
contribution of the methyl-group rotation is larger than the one of the
phenylene $\pi-flip$ in the 0.2-10 ps time interval. A broad maximum in the
VDOS was observed at 37 meV, that corresponds to the methyl-group
libration, i.e., to the attempt frequency for rotation of about $10^{-13}$
$s^{-1}$. At the highest temperature of observation, 390 K, the lowest time
of rotation, that is obtained from these  values of activation energy and
attempt frequency, is 10 ps. Consequently, the very short time $\tau_{1}$
($\simeq 0.3\,ps$), that is approximately temperature independent, cannot
correspond to the methyl-group rotation or phenylene $\pi-flip$. \par

The VFAM is relatively well characterized. Its decay time $\tau_{1}$ is
approximately independent of temperature $T$, and of momentum transfer $Q$
(Table-1). On the other hand the part $\rho$ of $S_{nh}(Q,t)$, due to the
VFAM, increases with $T$  and a little with $Q$ (Table-1). The weak
Q-dependence of $\rho$  comes probably from  a component of rotation in the
VFAM \cite{Bee88}. With regard to the temperature dependence, it is
interesting to compare $\rho$ to the volume fraction of the dynamical holes
determined by positron annihilation lifetime spectroscopy (PALS). From the
principle of the technique, the PALS lifetime of the  dynamic holes, which
have a size close to 0.3 nm, is longer than 2 ns \cite{Hri96}, i.e much
longer than $\tau_{1}$. It corresponds to the long lifetime $\tau_{3}$ of
the ortho-positronium (o-Ps). In consequence the lifetime of the dynamic
holes can be considered as infinite for the VFAM. In Figure-5 the ratio
$\rho$ is plotted against the dynamic hole volume fraction, $\Delta F_{h}$,
that was measured by PALS \cite{Hri96}. $\Delta F_{h}$ at a temperature T
was determined by subtracting the hole volume fraction, that was obtained
by extrapolation at $T=0 K$, from the total hole volume fraction at the
temperature T. Figure-5 shows a remarkable proportionality between $\rho$
and $\Delta F_{h}$ in the 170-390 K temperature interval. It is clear that
the temperature dependence of $\rho$ corresponds to that of the part of the
polymeric glass, which participates to the VFAM. \par

 A proportionality between $\Delta F_{h}$ and the QELS intensity of PMMA
and of PC,  respectively, was recently observed \cite{Duv98,Niv97}. From
the behavior of the VFAM similar to the one of the QELS, as compared to the
$\Delta F_{h}$, it would be deduced that the VFAM observed by neutron
scattering corresponds to the QELS. It does not seem to be the case,
because, principally, the time $\tau_{1}$ is more than three times shorter
than the one measured, and theoretically justified,  for the QELS
\cite{Niv97,Nov98}. However there is certainly a connection between the
VFAM and QELS, which can be found if we consider the interpretation of the
QELS \cite{Niv97,Nov98}, and the time resolved o-Ps annihilation
\cite{Vas99}. Novikov showed that the origin of the QELS can be the
third-order anharmonic term in the vibrational Hamiltonian \cite{Nov98}, so
that the QELS is proportional to the square of the Gr\"uneisen coefficient.
From  another point of view, the same author and collaborators \cite{Niv97}
showed theoretically that the QELS is proportional to $\Delta F_{h}$, by
adding to the harmonic potential a term of interaction between the
vibration and the fluctuation of the dynamic hole volume or of the free
volume. If the Hamiltonian in ref. \cite{Nov98} does not depend necessarily
on the presence of dynamic holes, the term of interaction, between the hole
volume fluctuation and the vibration,  originates from  third order terms
(enhanced by the dynamic holes) in the vibrational Hamiltonian. In the
model of Novikov \cite{Nov98}, the fact, that the  inverse of the
characteristic time  corresponds approximately to one third of the boson
peak frequency, is justified because  the anharmonic third order term is
relatively strong for the modes in the boson peak. \par

The inverse time, $1/\tau_{1}$,  corresponds to a frequency  of 0.5 THz
($\sim$2 meV) , that is very close to  the frequency of the boson peak
(1.65 meV) \cite{Sav99}, and not to one third. This shows that the  VFAM is
not identical to the QELS, and can not be interpreted simply by the
anharmonic third-order terms in the vibrational Hamiltonian,  and by the
model of Novikov \cite{Nov98}. The presence of a dynamic hole close to a
vibrating  polymeric nanodomain does not only damp, but can probably, in
given situations depending upon the hole volume, the vibration frequency or
the position of the hole relatively the vibrating domain, overdamp the
low-frequency vibrational modes. In the case of overdamping, the relation
between the decay time $\tau$ and the frequency $\nu$ of the vibration is
$2\pi\nu\tau=1$. With $\tau=\tau_{1}$, the frequency $\nu$ (0.5 THz) is
slightly higher than the one of the boson peak (0.4 THz). The suggested
interpretation is, therefore, that the VFAM corresponds to the
low-frequency modes in the boson peak, which are overdamped by nearby
dynamic holes. This is in agreement with the proportionality between $\rho$
and $\Delta F_{h}$  (Figure-5), and the Q-independence of $\tau_{1}$ or the
localization. This interpretation is confirmed by the time resolved o-Ps
annihilation \cite{Vas99}. Vass et al. showed that, for an organic glass
forming liquid, the thermal behavior of the ratio $I_{3}/I_{1}$ ($I_{1}$
and $I_{3}$ being  the intensities of respectively the fast p-Ps
and the slow o-Ps annihilations) can be understood if  very  fast molecular
motions (characteristic time less than 1 ps) make the dynamic hole volume
to fluctuate and, then, the  o-Ps, that is trapped in the hole, to
annihilate with a non negligible probability \cite{Vas99}.  Finally, the
dynamic holes or free volumes have two effects, on the one hand, they
enhance the effect of the anharmonic third order terms in the potential,
and they also possibly overdamp the vibration of neighboring molecular
groups. At this point, it must be noticed that very recently Schmidt et al.
\cite{Sch00} questioned the effect of the hole volume on the QELS, because
they observed that the QELS of PMMA did not depend very much on the
density, which varied from a sample to another by the pressure applied
during the cooling of the polymer. However, it was carefully verified that,
from T = 100 K up to $T_{g}$, the QELS  is proportional to the dynamic hole
volume, and not to the total hole volume \cite{Niv97,Duv98}. Now, it is
likely that the pressure applied during the cooling has principally an
effect on the static hole volume, on which the QELS does not depend. \par

The time $\tau_{1}$ is weakly dependent on temperature. However, its slight
decrease with temperature seems unavoidable for a good fit of the
experimental decays (Figure-3). If, a priori, it is surprising to observe a
relaxation time, that decreases with temperature, this thermal behavior of
$\tau_{1}$ is consistent with the proposed model of overdamping. As
recalled above, from a temperature of 100 K, there is a clear softening of
the low-frequency vibrational modes.The characteristic time $\tau_{1}$ is
inversly proportional to the vibration frequency $\nu$ ($\tau_{1}\approx
1/2\pi\nu$), and  therefore one expects that $\tau_{1}$ decreases with
temperature, as it is experimentally observed (Figure-3). The effect of the
dynamic holes on the fast dynamics,  which are the apparent result of a
slow relaxation, would deserve a more extensive study. It is likely that
the softening of the low-frequency modes is due to the presence of the
dynamical holes. \par

One can be surprised to find a so short relaxation time, or, in other
words, that vibrational modes of a frequency, that is equal to 0.5 THz, are
overdamped at $T<T_{g}$. This can be justified by the relatively large
volume fraction of dynamical holes in PC (Figure-5). On the other hand,
similar relaxation times were obtained in other glasses like the
ortho-terphenyl, respectively, by depolarized light scattering
\cite{Pat94}, and by Brillouin scattering \cite{Mona99}. \par

The limited resolution of the IN6 neutron spectrometer, in spite of its
deconvolution from the experimental dynamical structure factor, does not
allow to be confident of $S_{nh}(Q,t)$ at times longer than 10 ps, so that
the determined value of the $\tau_{2}$ relaxation time is rather imprecise.
However some properties of the slower relaxation, at which  all the glass
participates, contrary to the very fast relaxation, are clear enough:
$\tau_{2}$ is temperature and momentum transfer dependent (Table-1), like
the fast relaxation observed by Floudas et al. \cite{Flou93}. From
Figure-6, an activation energy of about 10.5 kJ/mole can be deduced.
$1/\tau_{2}$ increases clearly with Q, but the experimental precision does
not allow to give a quantitative Q-dependence. As a matter of fact,
$\tau_{2}$ is an effective relaxation time, which accounts certainly for
several different molecular motions. The activation energy for the
methyl-group rotation is 15-21 kJ/mole \cite{Schm85}, as noted above, and
14.6 kJ/mole for the segmental -C-C- rotational motion \cite{Hut91}. Both
activation energies are close to that deduced for $\tau_{2}$ (Figure-6), so
that it is likely that $\tau_{2}$ accounts for the methyl-group rotation
and the segmental relaxation. The Q-dependence of the effective time
$\tau_{2}$ may indicate that the relaxation is partially diffusion-like on
short molecular distances, and would correspond to the short-time regime of
the segmental relaxation \cite{Col98}. \par

\section{CONCLUSION}

The experimental  neutron scattering study brings new informations about
the anharmonic motions in a polymeric glass at $T<T_g$. The anharmonic
motions appear from a low temperature (50 K) in comparison with $T_{g}$
(420 K). A very fast relaxation characterized by a time close to 0.3 ps,
that is momentum transfer independent and decreases very slightly with
temperature, is observed at temperatures higher than 130K. This is
interpreted by the overdamping of the low-frequency vibrational modes, that
are responsible for the boson peak. The identical thermal behavior for the
proportion of the glass, that is submitted to the very fast anharmonic
motion, and for the dynamical hole volume fraction, shows that the
overdamping of the low-frequency modes is due to the presence of the
dynamical holes in the neighboring of the vibrating nanodomain. \par

\noindent ACKNOWLEDGEMENTS 

The authors are very grateful to K. Ngai and V.N. Novikov for valuable
discussions and suggestions.

\begin{figure}
\label{f1}
\caption{Mean square displacement $\langle u^2\rangle$ plotted against the
temperature.                           }
\end{figure}

\begin{figure}
\label{f2}
\caption{Debye-Waller corrected $S(Q,\omega)$ scattering spectra. (a) for
$Q=1.9\, \AA^{-1}$, at different temperatures: 15 K (full line), 30 K
(triangles), 110 K (circles), 220 K (squares), 300 K (stars). (b) at T =
390 K, for different momentum transfers: Q=0.91 $\AA^{-1}$ (squares), 1.33
$\AA^{-1}$ (stars), 1.9 $\AA^{-1}$.}
\end{figure}

\begin{figure}
\label{f3}
\caption{Normalized   scattering function from non-harmonic or relaxational
motions, for  $Q=1.9\, \AA^{-1}$, at different temperatures: 390 K (plus),
340 K (crosses), 300 K (stars), 260 K (open squares), 220 K (full squares),
170K (open circles). The decays are fitted with the expression \ref{exp},
using the parameters given in Table-1. The insert shows the decays, that
are obtained by the Fourier transform, from 0 ps to 2 ps in linear
coordinates, with the fits (full lines).}
\end{figure}

\begin{figure}
\label{f4}
\caption{Normalized  scattering function from non-harmonic or relaxational
motions,  at $T=390 K$, for different momentum transfers: Q=1.9 $\AA^{-1}$
(plus), 1.62 $\,\AA^{-1}$ (crosses), 1.33 $\,\AA^{-1}$ (stars), 1.11
$\,\AA^{-1}$ (open squares), 0.91 $\,\AA^{-1}$ (full squares)  . The decays
are fitted with the expression \ref{exp}, using the parameters given in
Table-1. The insert shows the decays from 0 ps to 2 ps in linear
coordinates.}
\end{figure}

\begin{figure}
\label{f5}
\caption{The ratio $\rho$ of the very fast relaxing part is plotted against
the dynamic hole volume fraction $\Delta F_{h}$. The temperatures
corresponding to the different points are: 170K, 220K, 260K, 300K, 340K,
390K.}
\end{figure}

\begin{figure}
\label{f6}
\caption{The inverse time $1/\tau_{2}$ is plotted against the inverse of
temperature for $Q=1.9 \AA^{-1}$. The dotted line corresponds to an
activation energy of 2.5 kcal/mole.}
\end{figure}

\begin{table*}
\caption
{Parameters in \ref{exp} deduced from the fits in Figures 3 and 4
}
\begin{tabular}{ccccc}
Q($\AA ^{-1})$ & T(K) & $\rho$& $\tau_1(ps)$ & $\tau_2(ps)$\\
\hline
1.9 & 170 &0.18& 0.22 & $\infty$ \\
1.9&220&0.25&0.24&634\\
1.9&260&0.31&0.25&287\\
1.9&300&0.37&0.28&163\\
1.9&340&0.43&0.30&100\\
1.9&390&0.51&0.33&60\\
1.62&390&0.43&0.33&87\\
1.33&390&0.32&0.33&129\\
1.1&390&0.27&0.31&149\\
0.91&390&0.24&0.27&162\\

\end{tabular}
\end{table*}

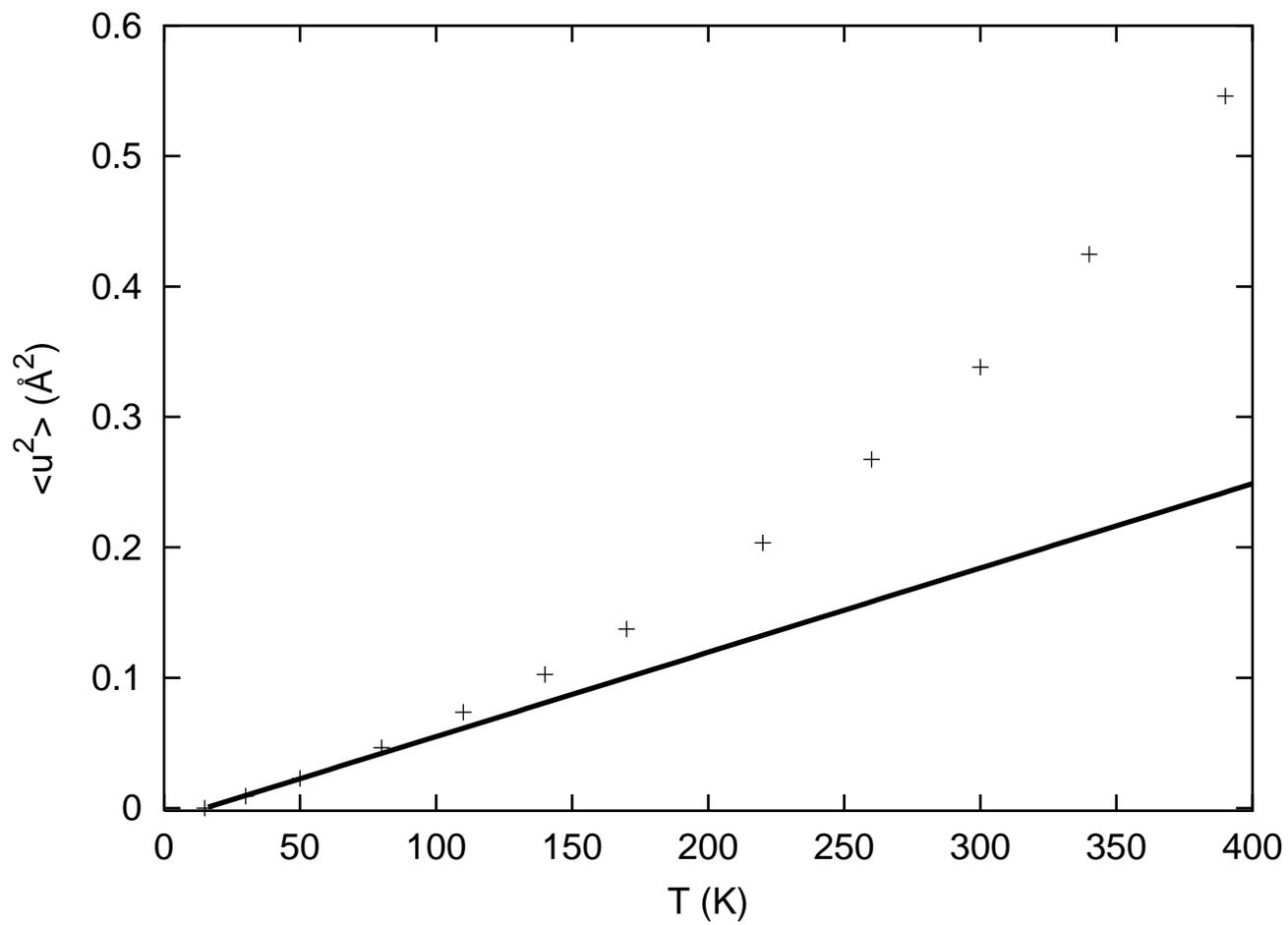
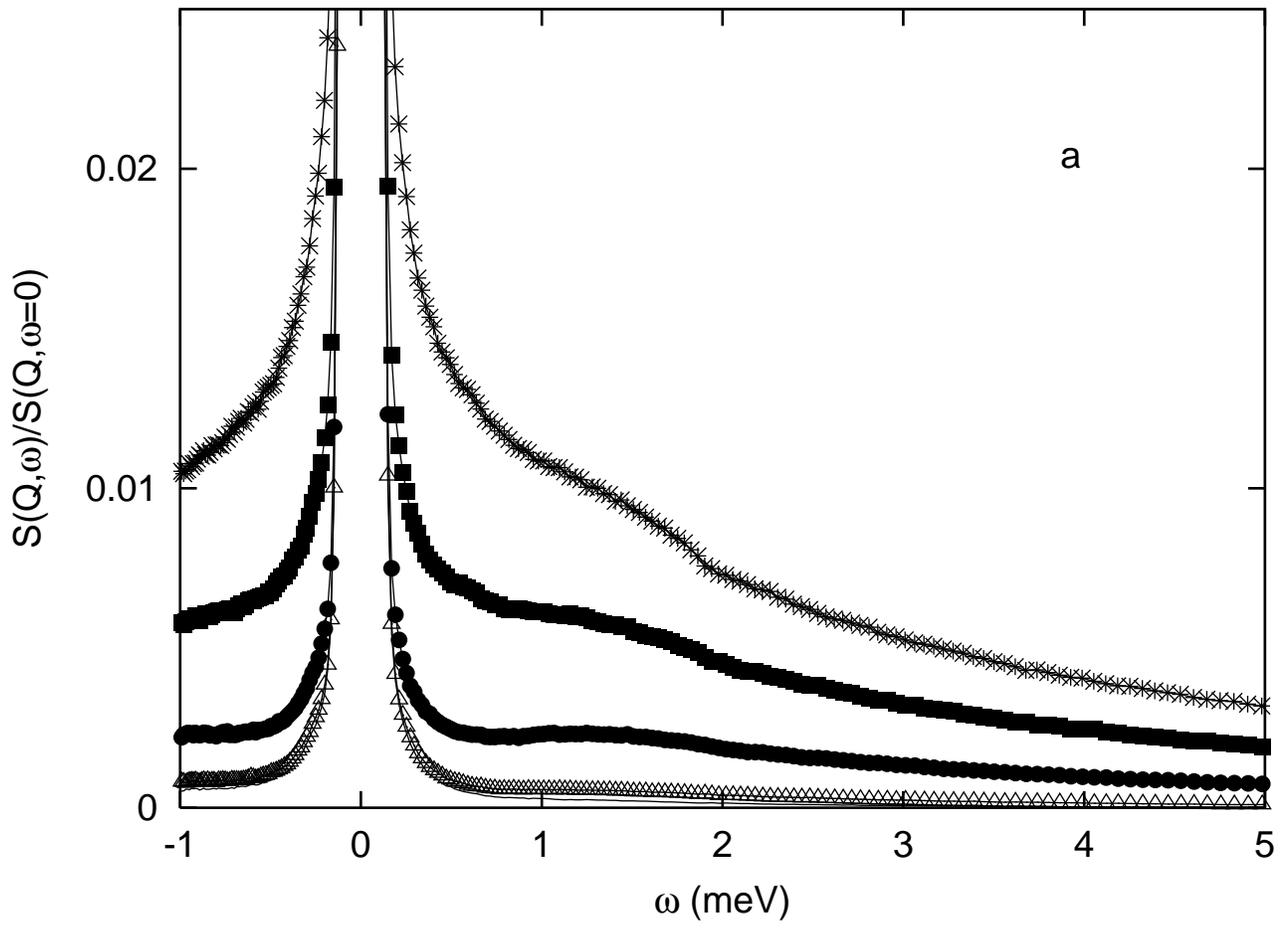
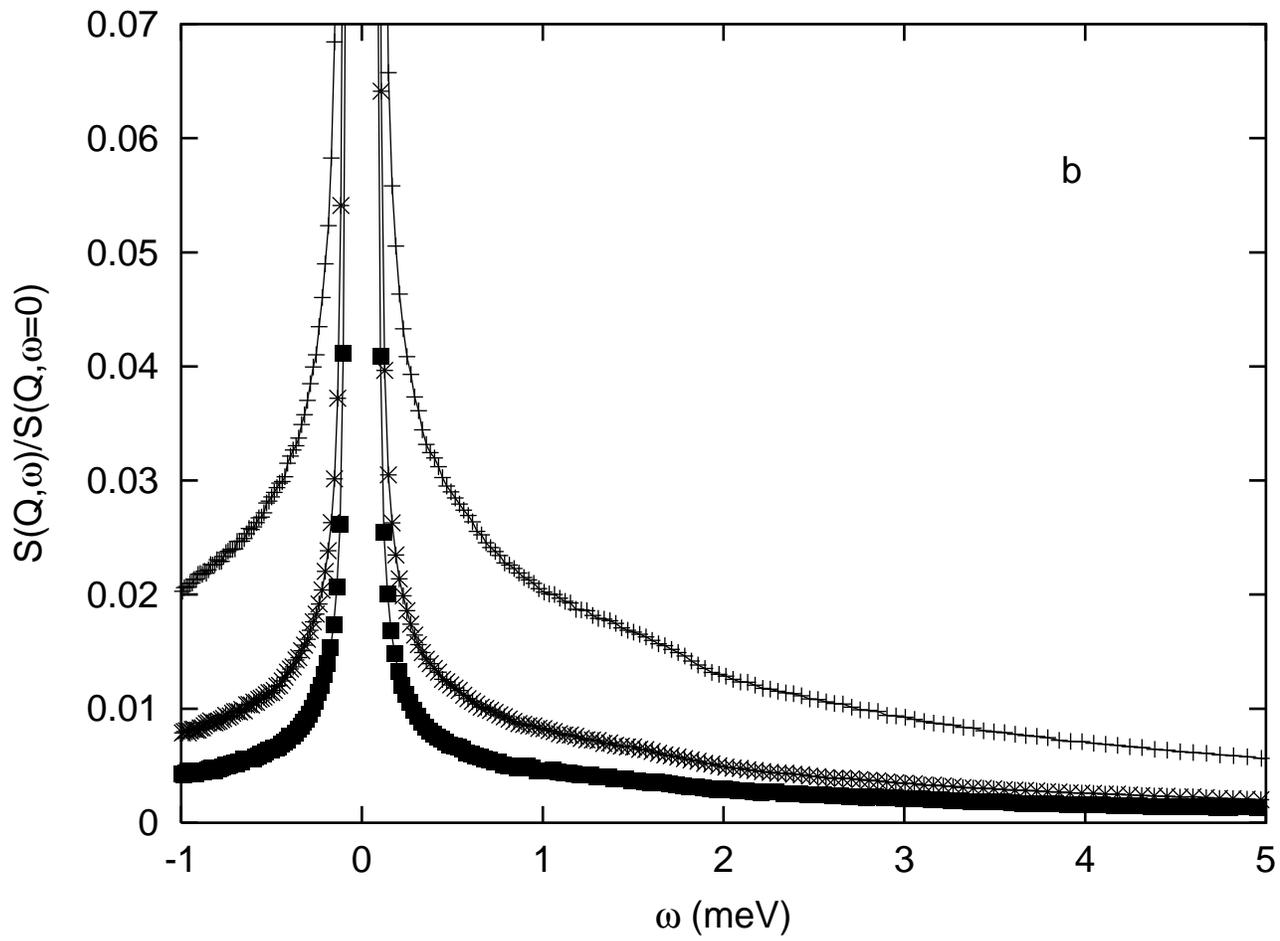
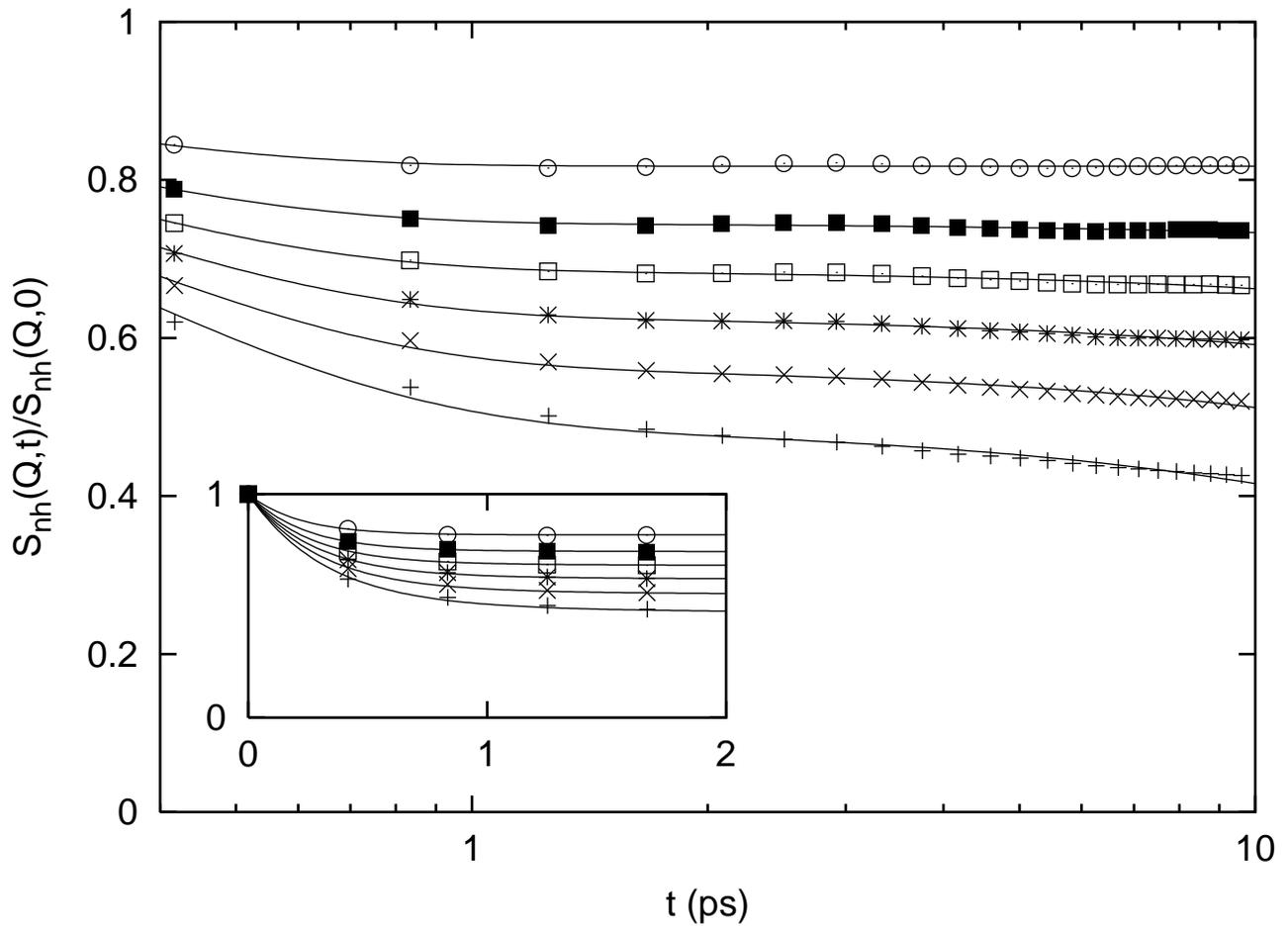
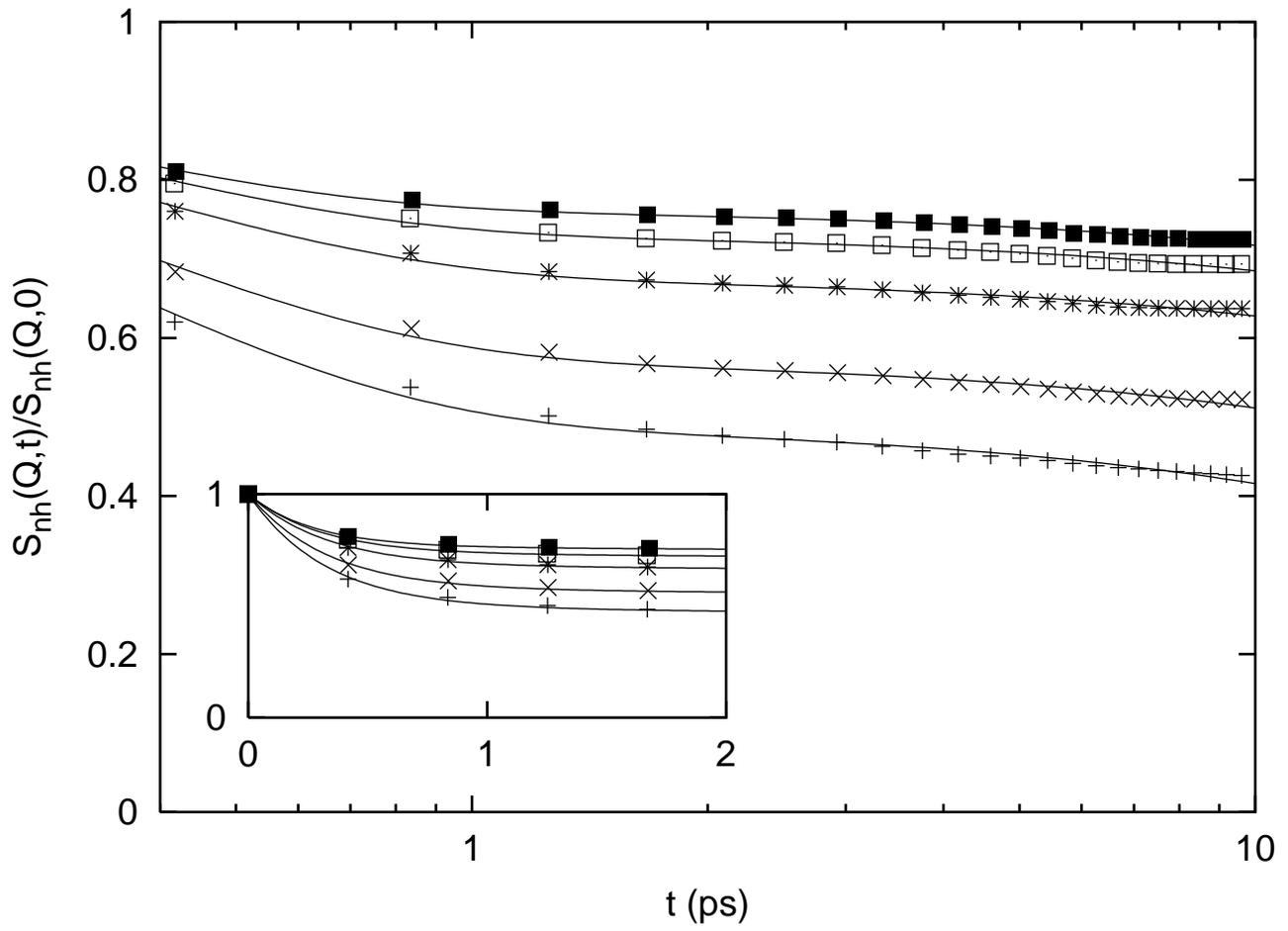
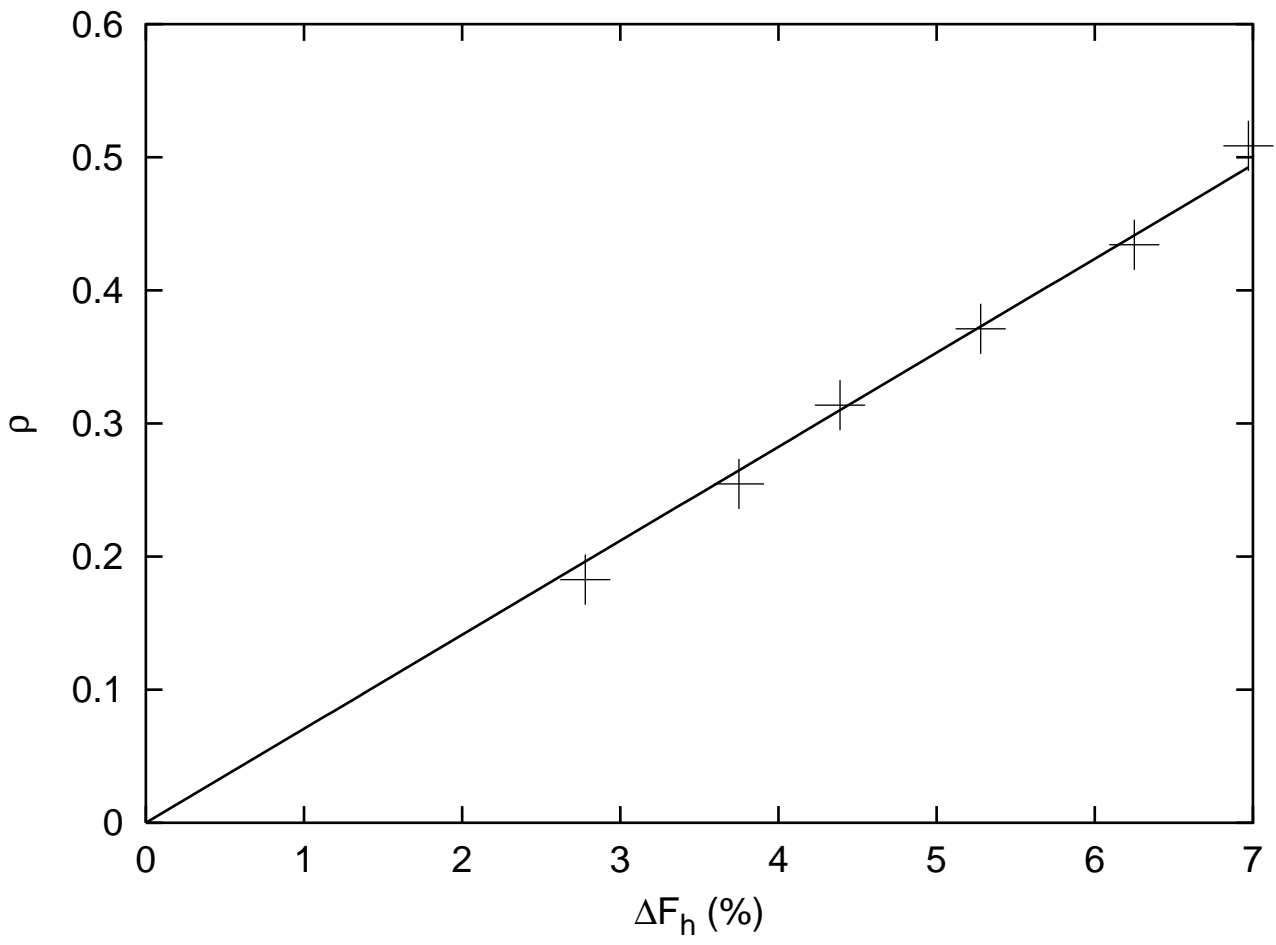
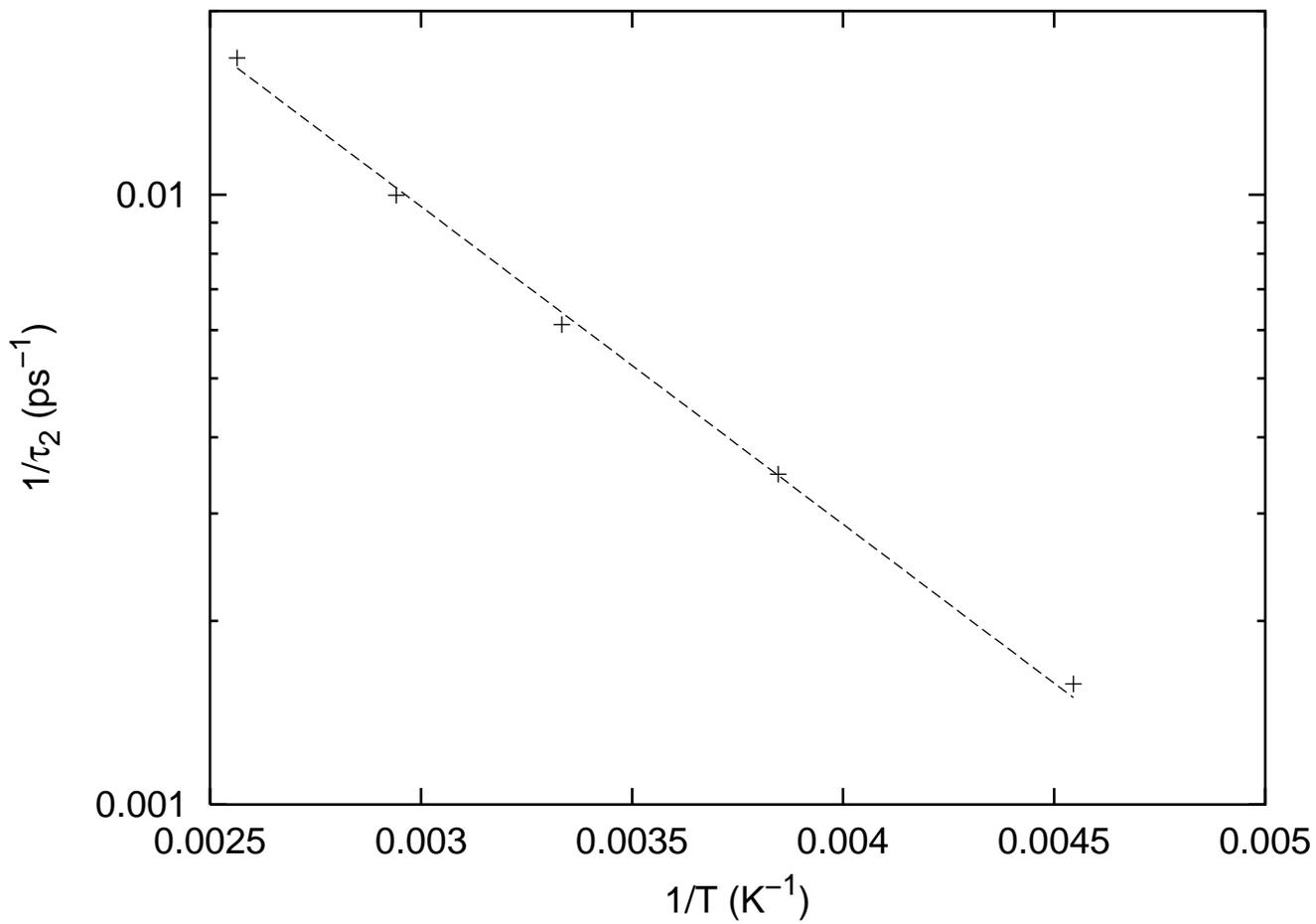
\end{document}